\documentclass[11pt]{article}
\usepackage{fleqn,cospar}

\usepackage{graphicx}
\input epsf.sty
\def \th {\thinspace}

\def\sun{\hbox{$\odot$}}
\def\approxgt{\mathrel{\hbox{\rlap{\lower.55ex \hbox {$\sim$}}
  \kern-.3em \raise.4ex \hbox{$>$}}}}
\def\approxlt{\mathrel{\hbox{\rlap{\lower.55ex \hbox {$\sim$}}
        \kern-.3em \raise.4ex \hbox{$<$}}}}

\title{THE EMISSION REGIONS IN X-RAY BINARIES:
 DIPPING AS A DIAGNOSTIC}

\author{M.J. Church
\address{School of Physics and Astronomy,
University of Birmingham, Edgbaston, Birmingham B15 2TT, UK}
\address{Institute of Astronomy, Jagiellonian University, ul. Orla
171, 30-244 Cracow, Poland}}

\begin{document}

\maketitle

\begin{abstract}
X-ray dipping in the black hole binary Cygnus\th X-1, the Galactic jet source
GRO\th J1655-40 and in low mass X-ray binaries is discussed. It is shown
that spectral analysis strongly constrains emission models. Measurement
of dip ingress/egress times allows the sizes of extended emission regions
to be determined, notably for the Accretion Disk Corona which is responsible for 
Comptonization in X-ray binaries. In LMXB, the radius of the ADC is shown to be 
between $\rm {\sim 1\times 10^9}$ and $\rm {\sim 5\times 10^{10}}$ cm, 
an appreciable fraction of the accretion disk radius. This is
inconsistent with Comptonization models requiring a localized
Comptonizing region, for example, in the immediate neighbourhood of 
the neutron star. Results from a survey of LMXB using {\it ASCA} and
{\it BeppoSAX} reveal an approximate equality between the height of
the blackbody emission region on the neutron star and the height of
the inner radiatively-supported disk, suggesting either that there is
a direct causal link, such as a radial accretion flow between the
inner disk edge and the star, or an indirect link, as in the case of
accretion flow creep on the surface of the neutron star as suggested
by Inogamov \& Sunyaev. Finally,
the survey shows that the blackbody cannot originate on the accretion
disk as the required inner radii in many sources are substantially
less than the neutron star radius.
\end{abstract}

\section*{X-RAY DIPPING}

X-ray dipping consists of reductions in intensity that are generally at the 
orbital period due to absorbing material in a bulge in the outer accretion disk or 
in blobbiness in a stellar wind. Dipping allows emission processes to
be determined unambiguously since the requirement to fit non-dip plus several dip
spectra, especially using high quality instruments on {\it ASCA}, {\it
RXTE}, {\it BeppoSAX}, {\it XMM} and {\it Chandra},
strongly contrains emission
models. Dipping provides information in 2 ways: via spectra
and via ingress/egress times which can give the size of emission regions,
and of particular interest is the size of Accretion Disk Coronae (ADC)
where Comptonization takes place. Dipping is relatively common, and in the 
following, dipping is discussed in the Black Hole 
Binary Cyg\th X-1, in the Galactic Jet Source GRO\th J1655-40
and in LMXB where $\sim $10\% of sources are dippers.

\section*{DIPPING IN CYGNUS\th X-1}

\begin{figure}
\includegraphics[width=90mm]{fig1}
\includegraphics[width=88mm]{xte_dips}
\caption{Left: GIS lightcurve for the {\it ASCA} observation of 1995
May 9 in two energy bands, together with the hardness ratio formed
from these (Ba\l uci\'nska-Church et al. 1997). 
Right: Distribution of X-ray dips with phase from the {\it RXTE} ASM.}
\end{figure}

Dipping in the Low State of Cyg\th X-1, studied by Kitamoto et al. (1984), 
Ebisawa et al. (1996) and Ba\l uci\'nska-Church et al. (1997, 2000a),
is usually of short duration,
lasting a few minutes, and results in an increase in hardness ratio
as shown in the {\it ASCA} data of Fig. 1 (left). 
The spectrum of this state is dominated by a hard component plus a
weak blackbody identified with the disk, plus a reflection component.
During dipping, spectral evolution can be described below 10 keV 
(where reflection contributes little) by an absorber
which progressively covers the extended ADC emission region with
spectral model: {\sc ag*pcf*cpl} where {\sc ag} represents Galactic absorption,
{\sc pcf} is the covering fraction and {\sc cpl} is a cut-off power law to
represent the Comptonized emission.

\subsection*{Distribution of Dipping with Orbital Phase}

An investigation of the variation of dipping with orbital phase
by Ba\l uci\'nska-Church et al. (2000a), 
revealed that there is a smooth variation peaking at phase zero
(superior conjunction of the black hole) as shown in Fig. 1 (right), plus
a second peak at phase $\sim $0.6 which is evidence for a stream
from the Companion. The smooth variation with phase was shown to
follow the variation of column density in the stellar wind of HDE\th
226868 with phase, although this
is difficult to calculate accurately because of several factors
including the variation of ionization state in three-dimensional space as
a function of distance from the black hole, and the suppression of
wind acceleration by X-ray illumination (Blondin 1994). The agreement indicates
that dipping is due to blobs in the stellar wind.

Dip duration times can be used to determine blob sizes, giving
$\rm {\sim 10^9}$ cm (Kitamoto 1984) and from {\it Rosat} (see below) 
$\approxgt $ $\rm {4\times 10^9}$ cm, which combined with
column densities in dipping gives blob densities $\rm
{\approx 10^{13}-10^{14}}$ cm$^{-3}$, $\rm {10^2-10^5}$ times more dense than in the wind,
from which it follows that there is a low ionization state in the blobs
and suggesting that recombination into the blobs in the X-ray shadow
will cause blob growth.

\subsection*{Size of the Accretion Disk Corona}

Dipping allows the size of the Comptonizing region to be measured under
conditions when the angular size of the absorber is greater than the
angular size of the largest emission region. In {\it Rosat} PSPC observations
(Ba\l uci\'nska-Church \& Church 1999),
dipping was close to 100\% deep indicating that this condition was
met. Then the dip ingress time $\rm {\Delta t}$, typically 5--10 s,
depends on the diameter of the major source region,
i.e. the ADC. The velocity of absorbing blobs has a Keplerian component
of $\sim $400 km $\rm {s^{-1}}$ at the radius of the Companion, plus
additional wind velocity $\sim $2100 km s$^{-1}$ (Herrero et al. 1995).
Thus the ADC radius $\rm {r_{ADC}}$ for an order of magnitude 
velocity of 2000 km s$^{-1}$ is given by $\rm {v_{blob}\;=\; 2\, r_{ADC}/\Delta t\,}$,
so that $\rm {r_{ADC}}$ is $\rm {5\times 10^{8}}$--
$\rm {1\times 10^{9}}$ cm.

In recent years, values of the Comptonization break energy $\rm {E_{CO}}$ in Cyg\th X-1
have become available, allowing constraints to be put on the
electron temperature in the Comptonizing region. From Fig. 1 of Dove
et al. (1997) a value of $\rm {E_{CO}}$ of $\sim $ 230 keV can be
derived. The cut-off energy must be between $\rm {kT_e}$ 
and 3$\rm {kT_e}$ for plasmas that are optically thin or optically thick to
electron scattering respectively. Particular Comptonization models
provide specific values of $\rm {kT_e}$: e.g. $\rm {kT_e}$ = 150 keV, $\tau $
$\sim $ 0.3 (Haardt et al. 1993); $\rm {kT_e}$ $\simeq $ 90 keV, $\tau $ 
$\sim $ 1.5 (Dove et al. 1997). We will assume that $\rm {kT_e}$ is between 77 and
230 keV and use these values to calculate the maximum radius of an ADC
in hydrostatic equilibrium $\rm {r_{max}}$ having\break $\rm {(GMm_p/r}$ $>$
$\rm {kT_{ADC}}$, (where $\rm {m_p}$ is the proton mass), i.e.

\[
\rm {r_{max} \simeq {1.6\times 10^{11}\; M_x\over T_{ADC}\; M_{\sun}}\;\; (cm) }
\]

\noindent where $\rm {T_{ADC}}$ is the ADC electron temperature in $\rm {10^7}$ K. From this it follows 
that the maximum hydrostatic ADC radius is between 0.6--1.8$\times 10^{10}$ cm for the range 
of $\rm {kT_e}$ above. Thus the measured radius of the ADC is much smaller than the 
hydrostatic value in contrast with ADCs in LMXB discussed later.
This implies strong differences between black hole and neutron star
systems, presumably due to the effects of the neutron star in forming
the ADC.

\section*{DIPPING IN THE GALACTIC JET SOURCE GRO\th J1655-40}

\begin{figure}[!h]        
\includegraphics[width=90mm]{jet}
\caption {Deadtime corrected PCA light curve of GRO\th J1655-40 in the band 2--30 keV
with 16-s binning.}
\end{figure}

\vskip - 90 mm \hskip 90 mm
\begin{minipage}[t]{86 mm}
Figure 2 shows part of the {\it RXTE} observation of 1997 Sept 27 -- Oct 1 in which 
very strong dipping can be seen. The non-dip emission in this source is dominated 
by a disk blackbody comprising 90\% of the 2--25 keV flux, with 10\% Comptonized emission
(Zhang et al. 1997)
plus an iron line with highly red and blue shifted wings at 5.9 and 7.3 keV
(Ba\l uci\'nska-Church \& Church 2000) which was the first detection of a disk line 
displaying a strong gravitational redshift in a Galactic source.
Similar lines have been detected in 4U\th 1630-47 (Cui et al. 2000)
and in XTE\th J1748-288 (Miller et al. 2000). Dipping 
has been investigated by Kuulkers et al. (1998, 2000) and by
Ba\l uci\'nska-Church ({\it these Proceedings}) who includes the disk
line.
\end{minipage}

\vskip 18 mm \noindent
Dip spectra are well-described by a model: {\sc pcf* ( dbb + gau ) + ab*pl},
where {\sc dbb} is a disk blackbody term.  
The requirement that the Comptonized emission ({\sc pl}) is modelled by a
simple absorber indicates that the emission is less extended than the
disk blackbody. The disk line intensity decreases (with very high
significance), and can be modelled adequately by including it
within the progressive covering term which is strong evidence 
that the line is real, and indicates that the line emission region is
similar in size to that of the disk blackbody. Dip ingress/egress times
show that the diameter of the disk blackbody emitter is $\rm {170-370}$
Schwarzschild radii, i.e. 3.4--7.4$\rm {\times 10^8}$ cm.

\section*{DIPPING IN LOW MASS X-RAY BINARIES}

\begin{figure}[!h]                             
\includegraphics[width=120mm]{merg4b}
\vskip .5 mm
\includegraphics[width=120mm]{flc}
\vskip -4 mm
\caption{Upper: Light curve of the complete {\it BeppoSAX}
observation of XB\th 1916-053 in the band 1.65--10 keV.
Lower: folded light curves in the bands 1--2 keV and 4--10 keV and the
hardness ratio formed from these based on a period of 3000s.}
\end{figure}

Firstly, examples will be shown of two dipping sources: XB\th 1916-053
and XB\th 1323-619.
XB\th 1916-953 is of particular interest as the shortest period dipping
LMXB with P$_{orb}$ = 50 min, implying the degenerate nature of the
Companion and because of the 1\% discrepancy between the
X-ray and optical periods (Grindlay et al. 1988; Chou et al. (2000). Figures 3 and 4 show light curves and folded light curves in two
energy bands from the 1997 observation with {\it BeppoSAX}
(Church et al. 1998a). It can be seen that dipping is close to 100\%
deep at all energies below 10 keV showing that the absorber has larger
angular extent than all source regions.
\begin{figure}[!h]                             
\includegraphics[width=88mm,height=62mm]{sinf3}
\includegraphics[width=88mm,height=62mm]{dip}
\vskip -3 mm
\caption{Left: Non-dip spectrum of XB\th 1916-053 from the {\it BeppoSAX}
narrow field instruments in the energy band 0.1--200 keV.
Right: Spectral evolution in dipping showing absorbed and
unabsorbed emission (see text).}
\end{figure}
Figure 4 (left) shows the non-dip broadband spectrum of the source in
{\it BeppoSAX} showing dramatically that the spectrum extends to above
200 keV with a Comptonization break energy $\rm {E_{CO}}$ of
$\rm {80\pm 10}$ keV (Church et al. 1998a; see also Bloser et al. 2000).  
Determination of the
properties of the Comptonizing region, i.e. the ADC, depends on obtaining 
$\rm {E_{CO}}$ which requires an energy response to at least 100 keV.
The high value of $\rm {E_{CO}}$ implies a high value of $\rm {kT_e}$. This
contrasts with earlier work on LMXB in the limited band of 1--10
keV where values of $\rm {E_{CO}}$ $\sim$ 3 keV were often (erroneously)
obtained due to other sources of curvature in the spectrum,
which would give incorrect values for Comptonization parameters.
Figure 4 (right) shows the corresponding dip spectra in the LECS and MECS
instruments revealing the apparently complex evolution in dipping, one part 
of the spectrum above 4 keV heavily absorbed while the part below 4 keV is unabsorbed.
Conventionally, such spectra were modelled by assuming two
components, one absorbed but the other not, the unabsorbed component
having a strongly decreasing normalization in dipping (e.g. Parmar et
al. 1986). However, this decrease was difficult to explain physically. 
These non-dip and dip spectra were well-fitted by a two component
model consisting of point-like blackbody emission identified with the
neutron star plus\break 
\begin{figure}[!h]                             
\vskip - 4mm
\includegraphics[width=80mm]{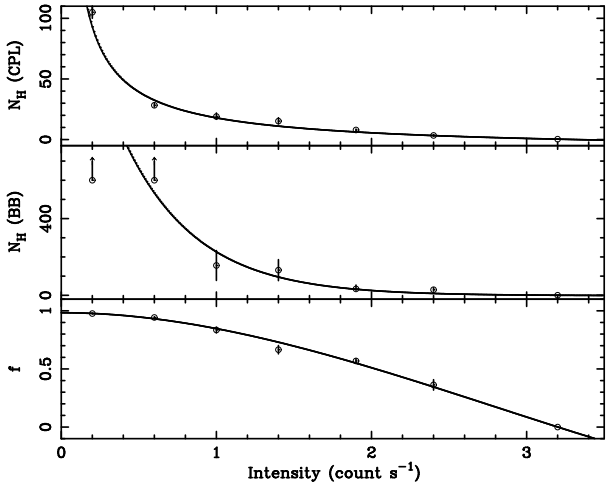}
\vskip -4mm
\vskip - 62 mm \hskip 90 mm
\begin{minipage}[t]{89 mm}
extended Comptonized emission from an ADC (Church et al. 1997), 
expressed as: I =
\[
\rm {I_0\, e^{-N_H^{Gal} \, \sigma} \,\{I_{BB}\, e^{-N_H^{BB} \sigma} +
I_{CPL}\,[\, f\,e^{-N_H^{CPL} \sigma} + (1-f)\, ] \} }
\]
where $\rm {N_H^{BB}}$ and $\rm {N_H^{CPL}}$ are the extra column
densities of the blackbody and cut-off power law in dipping,
and $\rm {N_H^{Gal}}$ is the constant interstellar column density.
This gave a simple explanation of spectral evolution during dipping. 
It is envisaged that a dense absorber
of large angular extent moves across the source regions
such that the extended ADC emission is progressively covered by
increasing amounts, the covering fraction
\end{minipage}
\vskip -1mm
\caption{Variation of $\rm {N_H}$ for blackbody and Comptonized emission
and progressive covering fraction in the {\it ASCA} observation of
XB\th 1916-053 (Church et al. 1997).}
\vskip -3 mm
\end{figure}

\noindent
rising smoothly from zero to 100\%
(Fig. 5). The point-like blackbody is at some stage almost instantly covered
and so rapidly removed. In this approach the unabsorbed part of the
spectrum is naturally explained as the uncovered ADC emission.
All dipping sources with an unabsorbed part of the spectrum have now
been explained in this way (Church et al. 1997, 1998b; Ba\l
ucinska-Church et al. 2000b; Smale et al. 2000). 
\begin{figure}[!hb]             
\vskip -4mm
\includegraphics[width=120mm,height=60 mm]{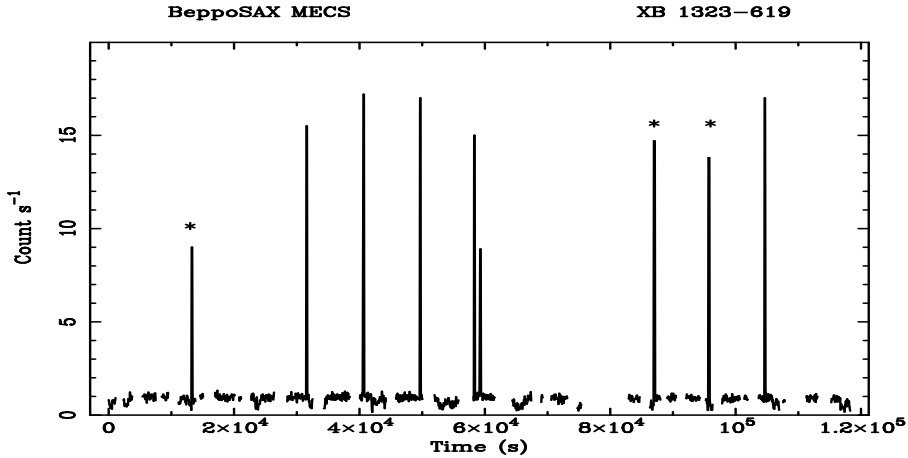}
\hskip 0mm\vskip -3mm
\caption{Background-subtracted MECS 2-10 keV lightcurve of XB\th
1323-619 from 1997 August 22.}
\vskip -2mm
\end{figure}

XB\th 1323-619 exhibits both dipping and quasi-periodic bursting. The
burst period has decreased from 5.30--5.43 hr during the {\it Exosat} observation 
(Parmar et al. 1989) to a value during the {\it BeppoSAX} observation of 
$\rm {\simeq 2.48\pm 0.08}$ hr, compared with an orbital (dip) period of 
$\rm {2.938\pm 0.020}$ hr (Ba\l uci\'nska-Church et al. 1999).
Thus bursts frequently occur during X-ray dips providing a unique diagnostic of the 
dramatic effects of a burst on the accretion disk.
The light curve shown in Fig. 6 contains several bursts in dips
indicated with an asterix. None of these shows the marked decrease in
intensity expected for the measured increase in $\rm {N_H}$ obtained
during dipping. Figure 7 compares spectral fitting of a normal burst
with a burst in a dip (peak
spectra). Attempts to fit the burst in a
dip with neutral absorber totally fail (left-hand panel), whereas
use of an ionized absorber
model provides a reasonable fit, thus
supporting the idea, previously suggested by Smale (1992) and Yoshida
(1993)
in the case of bursts in dips in XB\th 1916-053
that the outer disk, and thus all of the disk, becomes highly ionized
during a burst.\vfill \eject
\begin{figure}[!h]             
\includegraphics[width=90mm]{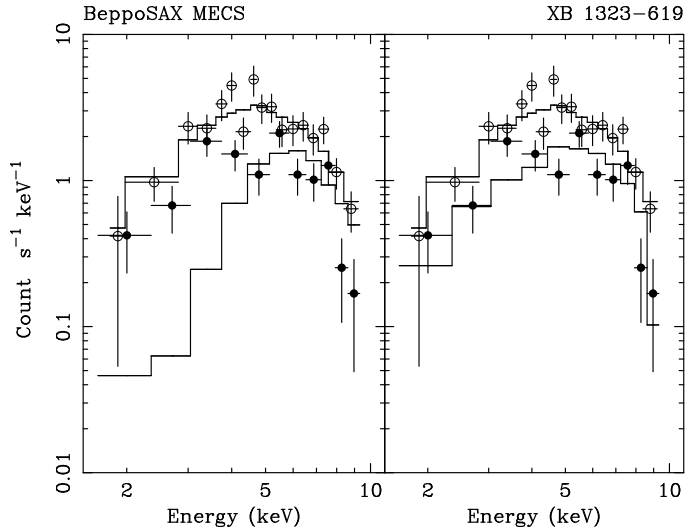}
\vskip - 70 mm \hskip 96.5 mm
\begin{minipage}[t]{83 mm}
\subsection*{Emission Models for Dipping LMXB}

Many different types of model have been used for LMXB including
one-component models with bremmstrahlung, blackbody, or Comptonized emission, and also
two-component models based on these. In the dipping sources, fitting is much more
constrained by the requirement to fit non-dip and several dip spectra, and it has been
shown that a blackbody plus Comptonization model is able to fit all of the sources
(Church \& Ba\l uci\'nska-Church 1993, 1995; Church et al. 1997,
1998a, Smale et al. 2000). More
recently the same model was shown to fit a range of Atoll and Z-track
sources (Church \& Ba\l uci\'nska-Church 2000).
These authors were able to show that the
\end{minipage}
\vskip - 4mm
\caption{MECS spectra of normal bursts (open circles) and of a burst
in a dip (filled circles). Left panel has neutral absorber; right panel
has ionized absorber for the burst in a dip.}
\end{figure}

\noindent
blackbody is point-like emission identified with the neutron star and that the 
Comptonized emission is extended as it is removed gradually in dipping and is 
identified with ADC emission. The size of this region may be measured
using dip ingress times in the dipping sources (below).

Although individual sources may be more complicated with line
emission also detected, and the ADC emission can vary 
in radius, $\rm {kT_e}$ and optical depth to electron scattering,
there is now strong evidence that these two basic components are present in {\it all}
LMXB (Church \& Ba\l uci\'nska-Church 2000), the fraction of blackbody however varying
between a very small fraction to $\sim $50\% of the total flux.
The question of whether the blackbody is from the neutron star or disk
which has been controversial is discussed later.

\subsection*{Deductions about Emission Regions: size of ADC}

Firstly, we summarise recent work in which dip ingress/egress times
are used to
derive the size of emission regions. For dipping which is 100\% deep 
(at any energy) the absorber must be of larger angular extent that the
emission regions, and so the ingress time $\Delta $t gives the diameter of the largest
source region, i.e. the ADC. The diameter $\rm {r_{ADC}}$ via:

\[
\rm {{2\pi \,r_{AD}\over P} =  {2\, r_{ADC}\over \Delta t}} 
\]

\noindent where $\rm {r_{AD}}$ is the accretion disk radius calculated using
standard theory, and P is the orbital period. Table 1 shows $\rm
{r_{ADC}}$ for several dipping LMXB obtained from recent work.
In the case of XB\th 1916-053,
the extreme mass ratio between the neutron star and the low mass
Companion makes the disk radius calculation uncertain, and the 
circularization radius is used as a lower limit.
The fraction of the disk  covered by the ADC is also shown, $\sim $15\% for several
sources but is as low as 2.5\% in the faint source XBT\th 0748-676 and as high 
as 52\% in the bright source X\th 1624-490.

\vskip - 5mm
\begin{table}[!h]
\vspace{-8mm}
\caption{Measured values of Accretion Disk Corona radius for 5
dipping sources.}
\begin{tabular}{lrrrrrl}
\hline
\
 & \hfil $\rm {\Delta t}$ (s) & $\rm {r_{AD}}$ (cm)& P
(hr)& $\rm {r_{ADC}}$ (cm)& $\rm {r_{ADC}/r_{AD}}$&ref.\\
\noalign{\smallskip\hrule\smallskip}
XB\th 1916-053 &130$\pm $20 & $\rm {2\times 10^{10}}$ &0.83 &$\rm
{2.1\times 10^{9}}$& 11\% &\scriptsize{Church et al. (1997)}\\
XB\th 1323-619  &250-600 & $\rm {3\times 10^{10}}$ &2.93 &2.2-5.4$\rm {\times 10^9}$  & 7.5-18\%
&\scriptsize{Ba\l uci\'nska-Church et al. (1999)}\\
X\th 1624-490 &12,500$\pm $2,500 & $\rm {1\times 10^{11}}$ &20.87  &$\rm {5.3\times 10^{10}}$ &52\%
&\scriptsize{Ba\l uci\'nska-Church et al. (2000b)} \\
XBT\th 0748-676 &110$\pm $30 & $\rm {3.4\times 10^{10}}$     &3.82   &$\rm {8.5\times 10^8}$ & 2.5\%
&\scriptsize{Church et al. (1998b)}\\
X\th 1755-338 &800$\pm $200 & $\rm {3.7\times 10^{10}}$      &4.46
&$\rm {5.8\times 10^9}$ &16\%&\scriptsize{Church \& Ba\l
uci\'nska-Church (1993)}\\
\hline
\end{tabular}
\hfil\hspace{\fill}
\end{table}

Next, spectral fitting results are used in Table 2 to constrain the electron temperature 
of the ADC, assumed constant. In the first three cases, {\it BeppoSAX} data allowed the
break energy $\rm {E_{CO}}$ to be derived with confidence.
\begin{table}[!h]
\vspace{-8mm}
\begin{minipage}{120mm}
\caption{Comparison of measured ADC radii with maximum radii of
hydrostatic ADC for sources having well-determined $\rm {E_{CO}}$.}
\begin{tabular}{lrrrr}
\hline
\
 & \hfil $\rm {E_{CO}}$ (keV) & $\rm {kT_e}$ (keV)& $\rm {r_{max}}$
(cm) & $\rm {r_{ADC}}$ (cm)\\
\noalign{\smallskip\hrule\smallskip}
XB\th 1916-053  &80& 24-80 & 2.4-7.9$\rm {\times 10^9}$& $\rm
{2.7\times 10^9}$\\
XB\th 1323-619 &44& 13-44  & 4.3-15$\rm {\times 10^9}$& 2.2-4.5$\rm
{\times 10^9}$\\
X\th 1624-490 &12& 4-12   & 16-53$\rm {\times 10^9}$& $\rm {53\times
10^9}$\\
\noalign{\smallskip}
\hline
\end{tabular}
\end{minipage}
\hfil\hspace{\fill}
\end{table}
$\rm {kT_e}$ is given as a range allowing for the extreme possibilities that the ADC is optically 
thin to electron scattering and $\rm {kT_e}$ = $\rm {E_{CO}}$ or is optically thick with 3$\rm
{kT_e}$ = $\rm {E_{CO}}$ (Haardt et al. 1993). The maximum radius of an ADC in hydrostatic equilibrium 
$\rm {r_{max}}$ is given by Eqn 1. Thus the values is Table 2 correspond to the high 
$\rm {kT_e}$ limit on the left for each source. Measured values of
$\rm {r_{ADC}}$ are included in the Table.

Comparison of $\rm {r_{max}}$ and $\rm {r_{ADC}}$ shows that for the first two relatively faint
sources XB\th 1916-053 and XB\th 1323-619, there is agreement between 
the measured $\rm {r_{ADC}}$ and the high-$\rm {kT_e}$ hydrostatic radius. 
For the bright source X\th 1624-490, there is also agreement, but with
the low-$\rm {kT_e}$ value suggesting that the ADC has higher optical depth
$\tau $. In this case, a value of $\tau $ $\sim $5.7 was derived
using the power law index of the spectrum, consistent with the above.
Thus, dipping ingress measurements show that the ADC covers a substantial 
fraction of the accretion disk, and in this limited sample, the ADC
radius agrees with the maximum hydrostatic radius beyond which
kT exceeds the gravitational force on an element of the medium 
and the ADC will be dissipated as a wind.

\section*{THE BLACKBODY SOURCE IN LMXB}

\begin{figure}[!h]                     
\includegraphics[width=90mm]{lums2}
\includegraphics[width=90mm]{hh1}
\caption{Left: Variation of blackbody luminosity with total 1--30 keV
luminosity for the survey of LMXB.
Right: Height of emitting region on the neutron star compared with $\rm
{H_{eq}}$, the equilbrium height of the inner radiative disk (see text).}
\end{figure}
\begin{figure}[!h]                     
\includegraphics[width=120mm,height=50mm]{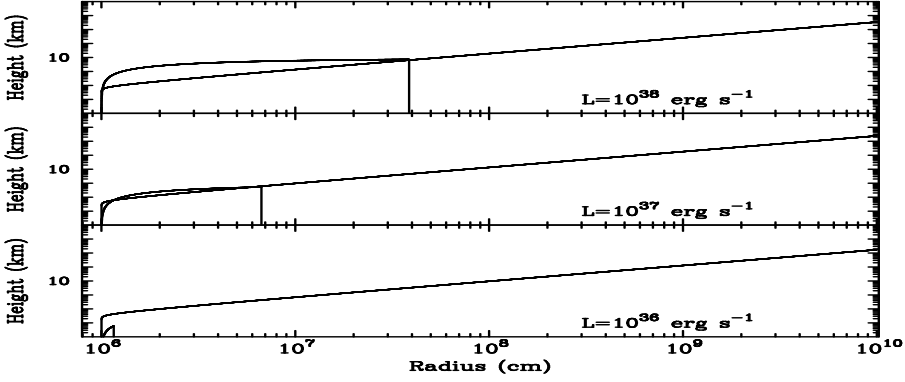}
\caption{Accretion disk height as a function of radial distance for 3
luminosities. The height of the Shakura-Sunyaev thin disk is shown
between radii of $\rm {10^6}$ and $\rm {10^{10}}$ cm, and on this is
superimposed the radiatively-supported disk (Eqn. 4).}
\end{figure}
In a survey of LMXB based on {\it ASCA} and {\it BeppoSAX} data, the
two-component model previously applied to dipping sources was shown to
give good descriptions of all of the LMXB sources tested
(Church \& Ba\l uci\'nska-Church 2000), and so is a good model for LMXB
in general. In Fig. 8 (left), the blackbody luminosity $\rm {L_{BB}}$ is
plotted against the total luminosity (1--30 keV) $\rm {L_{Tot}}$
and compared with the Newtonian value of 50\% of the accretion
luminosity (dotted line). It appears that in fainter sources there is
a tendency for $\rm {L_{BB}}$ to be a smaller fraction of the total.
These data were replotted in terms of the
emitting area on the neutron star assumed to be an equatorial strip
of half-height {\it h}. Then the emitting area of the star (a sphere
intersected by two parallel planes) is $\rm {4\, \pi \, R_x h}$
= $\rm {4\, \pi\, R_{BB}^2}$, where the blackbody radius $\rm {R_{BB}}$
is obtained from the measured flux. It is assumed that the inner part
of the accretion disk is radiatively supported, for $\rm {L_{Tot}}$ $>$
$\rm {3\times 10^{36}}$ erg $\rm {s^{-1}}$, with half-height {\it H}
increasing as a steep function of radius to a value $\rm {H_{eq}}$
as given in standard theory by:

\[
\rm {H = {3\, \sigma \, \dot M\over 8 \, \pi m_p\,c}
\left(1 - \left({R_x\over R}\right)^{1/2}\right) = H_{eq} \left(1 -
\left({R_x\over R}\right)^{1/2}\right)}
\]

\noindent
In Fig. 8 (right), h is plotted against $\rm {H_{eq}}$
calculated by obtaining $\rm {\dot M}$ from $\rm {L_{Tot}}$.
It can be seen that there is reasonable agreement between these
quantities except for lower luminosity sources. However, in these
cases, the radiative disk is unable to reach its full equilibrium
height. Figure 9 shows the radiative inner disk for three luminosities
demonstrating that only for $\rm {L_{Tot}}$ $\rm {\sim 10^{38}}$ erg
$\rm {s^{-1}}$ does the disk reach a stable value, i.e. $\rm {H_{eq}}$.

Because of this, the half-height of the disk was recalculated at the position
where the radiation pressure is 10 times the gas pressure $\rm
{H(r_{10})}$, as a good approximation to the radiative disk height (Czerny \& Elvis 1987).
The data were replotted in Fig. 10 (left), which showed good
agreement between h and $\rm {H(r_{10})}$ over a wide range of source
luminosity (Church \& Ba\l uci\'nska-Church (2000), as indicated 
schematically in Fig. 10 (right).
Two possible
explanations exist: i) the disk height {\it directly} determines h,
ii) H is a measure of $\rm {L_{Tot}}$ and some other physical process
determines h. Preliminary results indicate encouraging agreement 
between these results and the height of the emitting region on the
neutron star predicted by the theory of Inogamov \& Sunyaev (1999)
on the basis of accretion flow creep over the surface of the neutron
star. However, direct flow across the gap between the steep inner disk
edge and the star cannot be ruled out.
\begin{figure}[!h]                     
\includegraphics[width=90mm]{hh2}\vskip -55 mm \hskip 94 mm
\includegraphics[width=85mm]{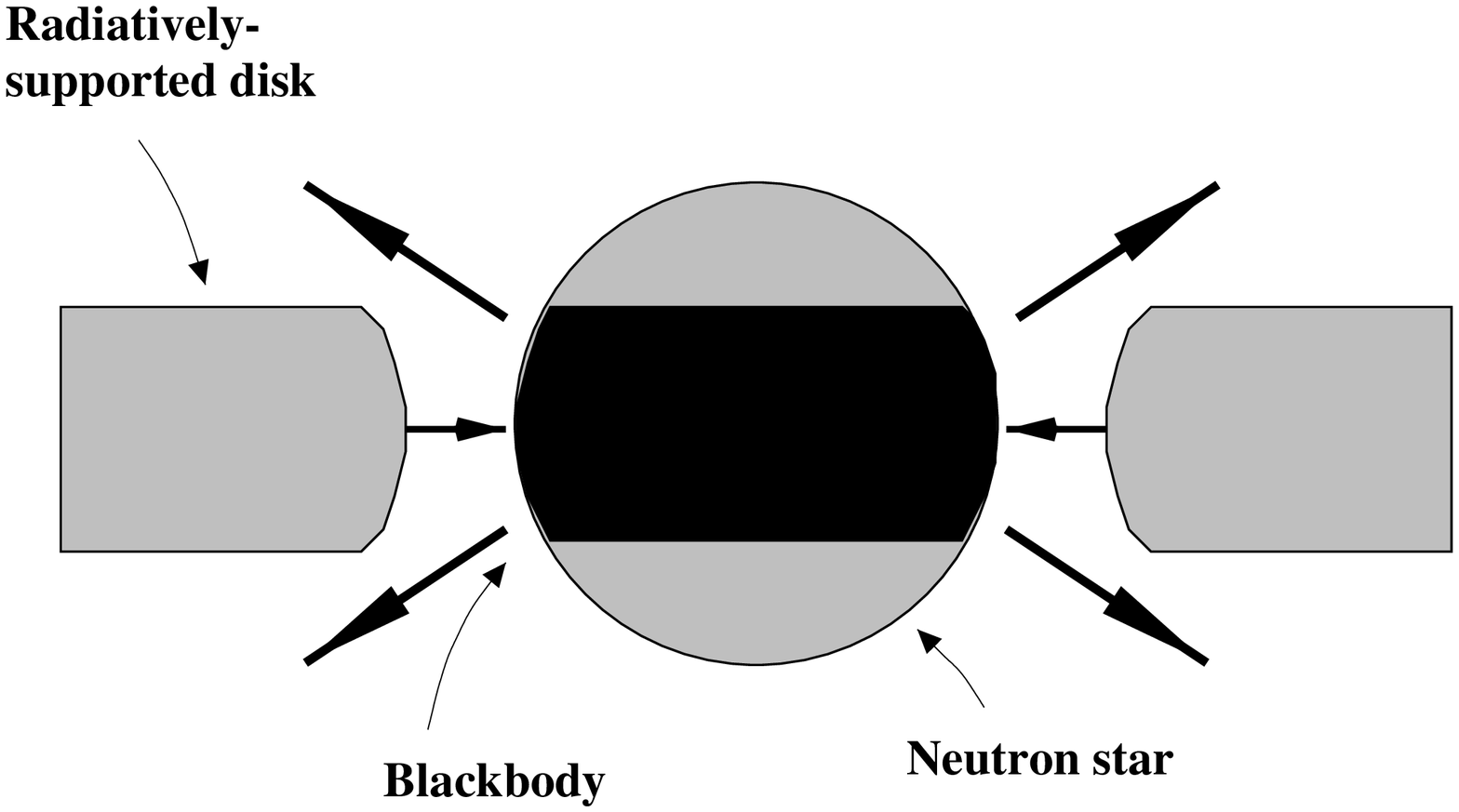}  
\vskip 10 mm
\caption{Left: Variation of h with $\rm {H(r_{10})}$, the height of the
radiative disk where $\rm {p_r}$ = 10$\rm {\;p_g}$ (see text). Right:
Schematic showing the equality h = H.}
\end{figure}

\subsection*{Application to Flaring in X\th 1624-490}

\begin{figure}[!h]                     
\includegraphics[width=150mm,height=60mm]{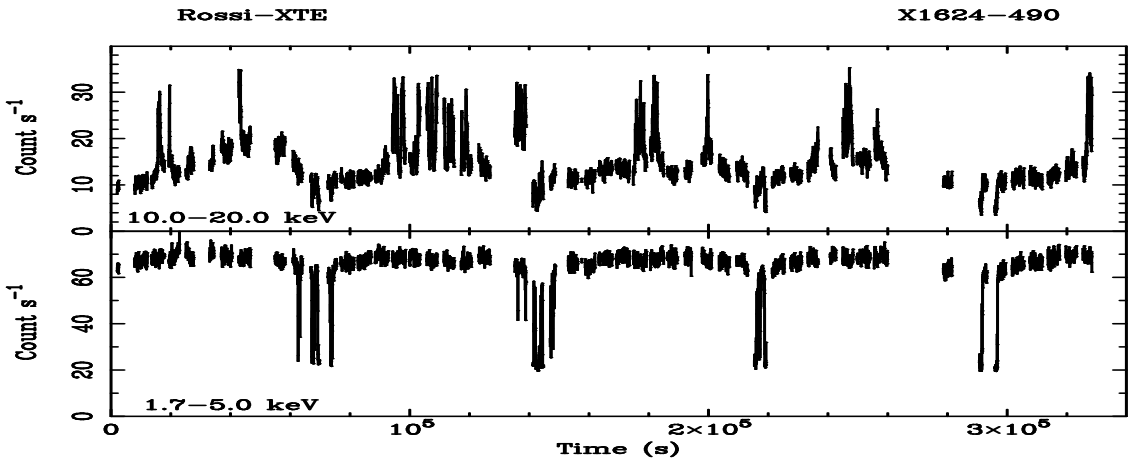}
\caption{Lightcurve of the 1999{\it Rossi-XTE}
observation of X\th 1624-490 in two bands showing the predominance of
dipping at low energies and of flaring at high energies.}
\end{figure}
If the agreement h = H is regarded as a model governing the
level of blackbody emission (although the mechanism 
is not established), there is already one situation in which
the model can explain the observed behaviour of a source, i.e.
flare evolution in X\th 1624-490. This flaring 
lasts for several thousand seconds, and in the 4-day observation
using {\it RXTE} in 1999, September, flaring was particularly strong
(Fig. 11, Smale et al. 2000; Ba\l uci\'nska-Church et al.
2000c). Spectral analysis shows that during this flaring, $\rm {kT_{BB}}$
increases as previously found (White et al. 1985),
but the blackbody radius $\rm {R_{BB}}$ decreases
substantially. In Fig. 12 (left), h is plotted against $\rm
{L_{Tot}}$, and $\rm {H_{eq}}$ shown by the solid curve.
\begin{figure}[!h]                                     
\includegraphics[width=90mm]{hh_final}\vskip -60 mm \hskip 102 mm
\includegraphics[width=70mm]{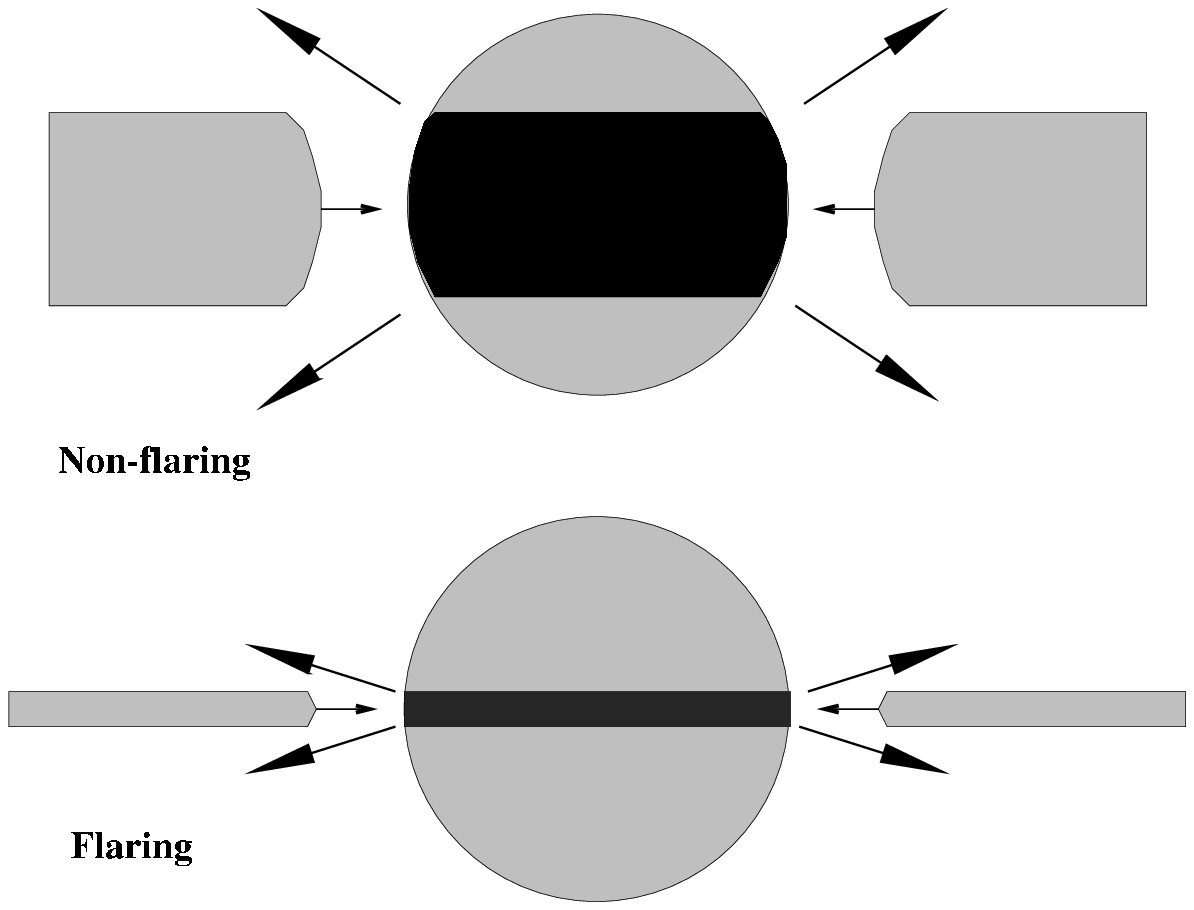}
\vskip 4mm
\caption{Left: Evolution of the blackbody emitter during flaring in X\th
1624-490 from observations with {\it BeppoSAX} and {\it RXTE} in 1997 and 1999
(Ba\l uci\'nska-Church et al. 2000).
Right: Schematic of the blackbody emission region in non-flare (top)
and flare peak.}
\vskip - 3mm
\end{figure}
If we assume that h = H also holds during flaring, then the radiative
disk height will be reduced in flaring (Fig. 12, right). A
possible explanation of this the high radiative flux of the blackbody
during flaring, with increased luminosity and reduced area, removes the
upper and lower parts of the inner accretion disk (Ba\l
uci\'nska-Church et al. 2000c). This reduces $\rm {\dot M}$ so
terminating the flare. This process constitutes a previously unknown
Eddington limit.

\subsection*{Neutron Star Blackbody or Disk Blackbody ?}

The results of the {\it ASCA} and {\it BeppoSAX} survey of LMXB
strongly indicates that the blackbody originates on the neutron star
since there would be no reason to expect a relation h = H if the
emission was from the disk. However, the sources included in the survey
were also analysed using a multi-colour disk (MCD) blackbody (Mitsuda et
al. 1984) plus Comptonization model with the results for inner radius 
$\rm {r_i}$ shown in Table 3. The ranges of inner radius values
include uncertainty in inclination angle and 90\% confidence fitting
uncertainties. 
\begin{table}[!h]
\vspace{-8mm}
\begin{minipage}{110mm}
\caption{Inner accretion disk radii $\rm {r_i}$ from fitting a disk blackbody + Comptonization model to
the sources in the {\it ASCA} and {\it BeppoSAX} survey of LMXB
(Church \& Ba\l uci\'nska-Church 2000).}
\begin{tabular}{llllll}
\hline
&& \hfil source & \hfil r$_i$ (km) &\hfil source &\hfil r$_i$ (km) \\
\noalign{\smallskip\hrule\smallskip}
&&Ser\th X-1             &2.7--4.7$^{+0.3}_{-0.6}$ &GX\th 9+9  &10--17$^{+3}_{-2}$\cr
&&X\th 2127+119          &0.79--1.1$^{+3.0}_{-1.1}$ &GX\th 13+1 &11--19$^{+10}_{-6}$\cr
&&Aql\th X-1             &2.8--4.8$^{+1.2}_{-0.8}$ &4U\th 1636-536 &14--24$^{+7}_{-4}$\cr
&&XB\th 1746-371         &1.0--1.4$^{+1.5}_{-1.1}$ &GX\th 5-1 &15--26$^{+4}_{-2}$\cr
&&XB\th 1254-690         &2.0--2.9$\pm $2.1        &Cyg\th X-2 &8.6--9.8$^{+1.0}_{-3.0}$\cr
&&XB\th 1916-063         &0.48--0.67$\pm $0.13\cr
&&XB\th 1323-619         &0.48--0.67$\pm $0.01\cr
&&X\th 1624-490          &0.39--0.41$^{+0.38}_{-0.21}$\cr
\noalign{\smallskip}
\hline
\end{tabular}
\end{minipage}
\hfil\hspace{\fill}
\vskip - 3 mm
\end{table}
In 8 cases, the inner disk radius of the disk blackbody is $<<$ than the neutron star radius; 
in other cases the power law flux increases with energy which is also unphysical.
It is known that 
the MCD model can underestimate the inner radius, however, the maximum correction 
factors that have been proposed are 2--4 (Merloni et al. 2000; Kubota et al. 2000), so that 
allowing for this, the very small values of $\rm {r_i}$ in Table 3 make it unlikely that the emission is from the disk, 
supporting the idea that disk photons are essentially completely reprocessed by Comptonization in the ADC.

\section*{CONCLUSIONS}

X-ray dipping can be regarded as a powerful diagnostic of the X-ray
emission regions as analysis of spectral evolution during dipping
strongly constrains emission models. Moreover, measurement of dip
ingress and egress times allows emission regions sized to be determined,
and in particular has revealed the very extended nature of the ADC
in LMXB with radii of $\rm {\sim 10^9}$--$\rm {5\times 10^{10}}$ cm,
suggesting the strong effect of the central object in both the
formation and heating of the ADC. These measurements exclude the
possibility that the Comptonizing region can be a localised region
such as in the vicinity of the neutron star. The recent survey of LMXB
based on {\it ASCA} and {\it BeppoSAX} data has demonstrated that
the height of the blackbody emitting region on the neutron star is
equal to the height of the radiatively supported inner disk. This result
will allow comparison with theory of flow between inner disk and star
and of the state and geometry of the inner disk. 
Finally, this survey has shown that the blackbody component 
cannot originate in the accretion disk.

\end{document}